\RequirePackage{fix-cm}
\documentclass[twocolumn,epjc3]{svjour3}
\RequirePackage[T1]{fontenc}
\smartqed  % flush right qed marks, e.g. at end of proof
\RequirePackage{graphicx}
\RequirePackage{mathptmx}      % use Times fonts if available on
\RequirePackage{flushend}
\RequirePackage{amssymb}
\RequirePackage{amsmath}
\RequirePackage{latexsym}
\RequirePackage[numbers,sort&compress]{natbib}
\journalname{}
\begin{document}

\title{Scalar-Tensor Teleparallel Gravity With Boundary Term by Noether Symmetries}


\author{Ganim Gecim\thanksref{e1,addr1},Yusuf Kucukakca\thanksref{e2,addr1}}

%placed here. General acknowledgments should be placed at the end of
%the article.
\thankstext{e1}{e-mail: gecimganim@gmail.com}
\thankstext{e2}{e-mail: ykucukakca@akdeniz.edu.tr}


\institute{Department of Physics, Faculty of Science, Akdeniz
University, 07058 Antalya, Turkey\label{addr1}}


\maketitle

\begin{abstract}
In the framework of teleparallel gravity, the Friedman-Robertson-Walker cosmological model with scalar tensor theory
where scalar field is non-minimally coupled to both the torsion scalar and boundary term is studied. Utilizing the Noether symmetry approach in such a theory, we obtain the explicit forms of the couplings and potential as a function the scalar field. We present some important cosmological solutions for the modified field equations using these functions getting via the Noether symmetry approach. Finally, the interesting cosmological properties of these solutions are discussed in detail, and it is shown that they can describe a universe lead to the late time accelerating expansion.
%\keywords{Hawking radiation\,tunnelling\, Warped AdS$_{3} black holes\, extremal black hle}
\end{abstract}
\section{Introduction}\label{intro}

Currently, one of the great problems of modern cosmology is to understand the late-time accelerated expansion of the universe. The idea of accelerated expansion of the universe has been confirmed by several astrophysical observations such as observations of supernovae Type Ia  (SNe Ia) \cite{riess99,Perll,Sperl}, cosmic microwave background radiation (CMB) \cite{netterfield02,bennet}, and large-scale structure \cite{tegmark04}. Despite all
these observations, this cosmological effect is not compatible with the existing equations of the standard Einstein's general theory of relativity. Therefore, the proposed solutions for explain the nature of the late-time accelerated expansion can be categorized into two classes in the literature. The first approach is to add an exotic fluid with a negative-pressure so-called dark energy to the matter part of the Einstein field equations. In this way, various dynamically dark energy models with different kind of scalar fields such as quintessence \cite{Rat,Zlatev}, phantom \cite{Cad}, quintom \cite{Guo}, fermionic fields \cite{Ar,Rib,Kremer,kuc1,kuc2,gec} and vector fields \cite{r1,r2} have been put forward to explain accelerating expanding of the universe. The second one is to modify the geometric part of the Einstein field equations at high energy levels. Hence the dark energy comes out from these modifications. Such a modifications in the framework of General Relativity (GR) are known $f(R)$ gravity in the metric \cite{Sot,fel,noji} and Palatini formalism \cite{pala,ama,olm,c1,c2}, $f(G)$ gravity \cite{odi,s1} and $f(R,G)$ gravity theory \cite{c3,c4,c5,c6,c7} where $f(R)$ and $f(G)$ is an arbitrary function of the curvature scalar and Gauss-Bonnet invariant respectively.

The teleparallel Gravity (TG) is an alternative gravitational theory to the GR. Although this theory gives the same field equations of the GR, the geometric formulations of these theories are different. On the one hand, GR is constructed from the curvature defined by the symmetric Levi-Civita connection that yields a vanishing torsion. Furthermore, TG considers a nonsymmetric Weitzenböck connection that has no curvature but only torsion. In other words, one can say that GR uses the curvature to geometrize the space-time, meanwhile teleparallel equivalent to GR uses torsion to explain the gravitational effects. It may be noted that in order to define the Weitzenböck connection in TG tetrad fields are used as dynamical objects whereas in order to define the Levi-Civita connection in GR, metric fields are used as dynamical variables. One of the interesting modified gravity model is the $f(T)$ gravity which is arbitrary function of the torsion scalar $T$. Many features of $f(T)$ gravity model have been considered in literature especially that in order to explain accelerated expansion of the universe through torsion \cite{ben,myr,pali1,pali2,yi} . Furthermore, some interesting black holes solutions have been found and studied, for example see \cite{pali3,na}.

Of particular interest is gravitational models allowing for nonminimal couplings between the scalar field and gravitational part. For example, one may consider a model of gravity with a non-minimally coupled scalar field to the curvature scalar in the form of $\xi R\phi^2$ where $\xi$ is a coupling constant and $R$ is the Ricci scalar \cite{m1,m2}. Such nonminimal couplings in the framework of GR have been studied in different contexts in the literature. It, naturally, appears when quantum corrections are considered and essential for the renormalizability of the scalar field theory in the curved space \cite{kit}. Moreover, this coupling has been used to explain both the early time inflation and late time cosmic acceleration of the universe in the metric and Palatini formalism \cite{q1,q2,q3,q4,q5,q6,q7}. On the other hand, an alternative gravitational scenario the so-called teleparallel dark energy, has been presented in the framework of TG in Ref. \cite{geng1,Wei1,ota1,sad,geng4,ss1,ja,ds,faz,sad1,t1} where authors considered the scalar field to be nonminimally coupled to the torsion in the form of $\xi T\phi^2$. Considering the nonminimal coupling between the scalar field and the torsion scalar opens a new window in analysing the cosmological evolution of the universe. It has been shown that this model has a richer structure, exhibiting quintessence-like or phantom-like behavior, or experiencing the phantom-divide crossing. Other studies for the model discuss parameter fit with cosmological observations \cite{geng2,Gu}, phase-space analysis \cite{Xu1}, Noether symmetry approach \cite{kucuk,srf}, growth of density perturbations \cite{gng}, the possibility of singularities \cite{geng3} and spherically symmetric solutions \cite{ss}. We note that there are some interesting models including fermionic and tachyonic field nonminimal coupled to the torsion discussing the cosmic evolution of the universe \cite{kuc1,gec,Bani,ota2,faz2,mot}.

Recently, Bahamonde and Wright \cite{ba1} proposed a different model in the teleparallel gravity framework by introducing a scalar field non-minimally coupled to the torsion scalar $T$ as well as to the boundary term $B$, corresponding to the divergence of torsion vector $B=\frac{2}{e}\partial_{\mu}(eT^{\mu})$. The model reduce to both nonminimally coupled teleparallel gravity and nonminimally coupled general relativity under certain limits. There it was found that this model generically yields to a late time accelerating attractor solution without requiring any fine tuning of the parameters. It was studied the parameterized post-Newtonian approximation \cite{sa} as well as thermodynamics aspects \cite{ba2} of this model. Using the Noether symmetry technique for this model, new wormhole solutions according to the Morris and Thorne paradigm has been derived \cite{ba3}. In Ref \cite{t2}, the behavior and stability of the scaling solutions are studied for scalar fields endowed with inverse power law potentials and with exponential potentials for this models.

One of the most popular methods of finding the exact solutions to the nonlinear higher-order differential equations is to use the Noether symmetry approach. Noether symmetry is associated with field equations possessing the lagrangian and it guarantees the existence of conserved quantities that allow to reduce dynamics thanks to the presence of cyclic variables \cite{rit1,rit2,cap1,cap2}. Moreover, the existence of this symmetry leads to a specific form of the unknown functions that appear in the Lagrangian. The method is used to obtain cosmological models in several alternative theory of gravity, for example, scalar tensor theory \cite{san2,camci,Basilakos,kuc3,muh,Kremer4,pali,isil,beli,beli1,o1,papag}, $f(R)$ theory \cite{Cap08,Vakili08,ros,pal,hus,sha,daraa,hy,c1,c2}, $f(T)$ theory \cite{wei2,ata,jam1,moh,hann,aslam,pal2,pal3}, the theories of gravity with a fermionic field \cite{Kremer,kuc1,kuc2,gec} and others \cite{k1,k2,k3,k4,k5,k6,k7,k8,k9}.  With the help of Noether symmetry, some cosmological analytical solutions for the scalar tensor TG were obtained in Ref \cite{kucuk,k10}. Having above points in mind, the main goal of this paper is to explore Noether symmetry in scalar tensor theory of gravity in which the scalar field is non-minimally coupled to both torsion scalar and the boundary term. We have determined the interesting physical forms of the coupling functions and potential by existence of Noether symmetry and found exact solutions of the field equations to discuss cosmic evolution of the universe via cosmological parameters. This paper is organized as follows. In Section \ref{sec:1}, we give a basic formulation of the scalar tensor teleparallel gravity including a boundary term. In Section \ref{sec:2}, we discuss the Noether symmetries of the flat Friedmann-Robertson-Walker space-time in the context of the considering model. We search the cosmological solutions by using the obtained forms of the coupling functions and potential. Finally, in Section \ref{sec:3}, we conclude with a brief summary.

\section{Teleparallel Theory With a Non-minimal Coupling to a Boundary Term}\label{sec:1}

We will now consider the following action which describes a non-minimally coupled scalar field to both torsion scalar and the boundary term \cite{ba1,ba3}
\begin{equation}
S=\int{ d^{4}x e \Big[\frac{1}{2\kappa^2}\Big(f(\phi)T+g(\phi)B\Big) + \frac{1}{2}
\partial_{\mu}\phi\partial^{\mu}\phi -
V(\phi)\Big]}. \label{action1}
\end{equation}
Here the boundary term is defined in terms of divergence of torsion vector $B=\frac{2}{e}\partial_{\mu}(eT^{\mu})$ where $e = det(e_{\mu}^a)=\sqrt{-g}$ is  the volume element of the metric, $V(\phi)$ is the scalar field potential and  $f(\phi)$ and  $g(\phi)$ are the generic functions describing the coupling between the scalar field and torsion scalar and boundary term, respectively. In fact, such a coupling of the scalar field to a torsion scalar and boundary term is not a new idea. For example, one can choose the coupling functions as $f(\phi)$$=$$1-\xi\phi^2$ and $g(\phi)=\chi\phi^2$ where $\xi$ and $\chi$ are coupling constants. For $\chi=0$, one can recover an action named the teleparallel dark energy model in which scalar field coupled to the torsion scalar. For $\xi=0$, it may correspond to gravity including boundary term. When one sets $\xi=-\chi$ one will recover an action which is scalar field models non-minimally coupled to the Ricci scalar. The minimally coupled quintessence theories arise when we take $\xi=\chi=0$. Additionally, teleparallel scalar tensor theory without the boundary term can be recovered, if we set $g(\phi)=0$ in the action (\ref{action1}). We note that any possible coupling of the scalar field with ordinary matter Lagrangian is disregarded and we use $\kappa^2=1$.

By varying the above action with respect to the tetrad field, one obtain the following gravitational field equations \cite{ba1,ba3}
\begin{eqnarray}
2f(\phi)\Big[e^{-1}\partial_{\mu}(eS_{a}{}^{\mu\nu})-e_{a}^{\lambda}T^{\rho}{}_{\mu\lambda}S_{\rho}{}^{\nu\mu}-\frac{1}{4}e^{\nu}_{a}T\Big] \nonumber \\+e^{\nu}_a \Big[\frac{1}{2}\partial_\mu \phi \partial^\mu \phi -V(\phi)\Big]
-e^{\mu}_a \partial^\nu \phi \partial_\mu \phi+e^{\nu}_{a} \Box g(\phi) \nonumber\\+2\Big[\partial_{\mu}f(\phi)+\partial_{\mu}g(\phi)\Big]e^\rho_a S_{\rho}{}^{\mu\nu}-e^\mu_a \nabla^{\nu}\nabla_{\mu}g(\phi)&=&0
 \label{fe}
\end{eqnarray}
where $\Box= \nabla_{\alpha}\nabla^{\alpha}$; $\nabla_{\alpha}$ is the covariant derivative with respect to the Levi-Civita connection. On the other hand, the variation of the action (\ref{action1}) with respect to the scalar field $\phi$ gives rise to the Klein-Gordon equation governing the dynamics of the scalar field
\begin{eqnarray}
\Box\phi+V'(\phi)-f'(\phi)T-g'(\phi)B=0.\label{KG}
\end{eqnarray}
where the prime indicates the derivative with respect to $\phi$.

Now, we consider the four-dimensional spatially flat Friedmann Robertson Walker (FRW) space-time with the
metric
\begin{equation}\label{FRW}
ds^2=dt^2-a^2(t)(dx^2+dy^2+dz^2),
\end{equation}
where $a(t)$ is the scale factor of the universe. The corresponding tetrad components for the FRW metric are  $e^i_\mu=(1,a(t),a(t),a(t))$. With this definition of the tetrad field, the torsion scalar and the boundary term can be expressed in terms of the scale factor and its time derivatives as follows \cite{ba1,ba2,ba4}
\begin{equation}\label{tb}
T=-6\frac{\dot{a}^2}{a^2},\quad B=-6\left(\frac{\ddot{a}}{a}+2\frac{\dot{a}^2}{a^2}\right),
\end{equation}
where dot denotes differentiation with respect to the time coordinate. Inserting the tetrad components for the flat FRW metric into the action (\ref{action1}) and using the equations (\ref{tb}) we find a point-like Lagrangian as follows,
\begin{equation}\label{lag}
\mathcal{L}=-3fa\dot{a}^2+3g'a^2\dot{a}\dot{\phi}+a^3(\frac{\dot{\phi}^2}{2}-V(\phi)).
\end{equation}
From the Euler-Lagrange equation for the scale factor $a$ applied to the above Lagrangian, we obtain the following acceleration equation
\begin{equation}\label{dec}
2\dot{H}+3H^2=-\frac{p_{\phi}}{f}.
\end{equation}
Here, $H=\dot{a}/a$ is the Hubble parameter. The modified Friedmann equation is obtained by imposing that the energy
function $E_{L}$ associated with the Lagrangian (\ref{lag}) vanishes, i.e.
\begin{equation}\label{fried}
E_{L}=\dot{a}\frac{\partial{L}}{\partial{\dot{a}}}+\dot{\phi}\frac{\partial{L}}{\partial{\dot{\phi}}}=0 \Rightarrow H^2=\frac{\rho_{\phi}}{3f},
\end{equation}
In the equations (\ref{dec}) and (\ref{fried}), the energy density and the pressure of the scalar field $\rho_{\phi}$ and $p_{\phi}$ are respectively defined as follow
\begin{equation}\label{ed}
\rho_{\phi}=\frac{1}{2}\dot{\phi}^2+V+3g'H\dot{\phi},
\end{equation}
\begin{equation}\label{pres}
p_{\phi}=\frac{1}{2}\dot{\phi}^2-V+2f'H\dot{\phi}-g'\ddot{\phi}-g''\dot{\phi}^2.
\end{equation}
These two expressions define an effective equation of state parameter $\omega_{\phi}=p_{\phi}/\rho_{\phi}$, which drives the behavior of
the cosmological model. Finally, from the Euler-Lagrange equation for the scalar field $\phi$ by using the Lagrangian (\ref{lag}), the modified Klein-Gordon equation takes the form
\begin{equation}\label{kg}
\ddot{\phi}+3H\dot{\phi}=\frac{1}{2}\left(f'T+g'B\right)-V.
\end{equation}
%\We see that the coupling functions contribute
It is clear that to obtain some cosmological solutions to the modified field equations, first of all one has to determines for a form of the potential
function $V(\phi)$ and the coupling functions $f(\phi)$ and $g(\phi)$ of the scalar field. In the next section, we will fix this issue by demanding
that the point-like Lagrangian of the action (\ref{action1}) satisfies the Noether symmetry condition.

\section{Noether Symmetry Approach and Cosmological Solutions}\label{sec:2}

In theoretical physics, it is important to develop techniques to find  the solutions of non-linear equations system. Noether symmetry approach has become an important technique to solve such a system. This approach provides an systematic way to find conserved quantities for a given Lagrangian. At the same time, as a physical criterion, this approach also allows one to select the unknown functions in gravity models.
%In this way, constraints on dynamics are achieved and it is possible to solve the equations of motion.

Now, we seek Noether symmetries for the Lagrangian (\ref{lag}). The Noether symmetry generator is a vector field on the tangent
space $\mathcal{T Q}=(a,\phi,\dot a,\dot\phi)$ defined by
\begin{equation}\label{vec}
\textbf{X}=\alpha\frac{\partial }{\partial a}+\beta\frac{\partial}{\partial \phi}+
\dot \alpha\frac{\partial }{\partial \dot a}+\dot \beta\frac{\partial }{\partial \dot\phi},
\end{equation}
where $\alpha$ and $\beta$ are both functions of the generalized coordinates $a$ and $\phi$. The Noether symmetry then implies
that the Lie derivative of the Lagrangian with respect to this vector field vanishes, that is, $L_\textbf{X}\mathcal{L}=0$,
which leads
\begin{equation}\label{noet}
L_\textbf{X}\mathcal{L}=\textbf{X}\mathcal{L}=\alpha\frac{\partial \mathcal{L}}{\partial a}+\beta\frac{\partial\mathcal{ L}}{\partial \phi}+
\dot \alpha\frac{\partial \mathcal{L}}{\partial \dot a}+\dot \beta\frac{\partial \mathcal{L}}{\partial \dot\phi}=0.
\end{equation}
In general, the Noether symmetry condition leads to an expression of second degree in the velocities ($\dot{a}$ and $\dot{\phi}$) with coefficients being partial derivatives of $\alpha$ and $\beta$ with respect to the variables $a$ and $\phi$. Thus, the resulting expression is identically equal to zero if and
only if these coefficients are zero. This gives us a set of partial differential equations for $\alpha$ and $\beta$. For the Lagrangian (\ref{lag}), the Noether symmetry condition (\ref{noet}) yields the following system of partial differential equations
\begin{equation}\label{n1}
f\left(\alpha+2a\frac{\partial\alpha}{\partial a}\right)+f'a\beta-g'a^2\frac{\partial\beta}{\partial a}=0,
\end{equation}
\begin{equation}\label{n2}
g'\left(2\alpha+a\frac{\partial\alpha}{\partial a}+a\frac{\partial\beta}{\partial \phi}\right)+g''a\beta-2f\frac{\partial\alpha}{\partial \phi}+\frac{a^2}{3}\frac{\partial\beta}{\partial a}=0,
\end{equation}
\begin{equation}\label{n3}
\alpha+2g'\frac{\partial\alpha}{\partial \phi}+\frac{2a}{3}\frac{\partial\beta}{\partial \phi}=0,
\end{equation}
\begin{equation}\label{n4}
3V\alpha+aV'\beta=0.
\end{equation}
We solve this system of equations to find the values of $\alpha$, $\beta$, $f(\phi)$, $g(\phi)$ and $V(\phi)$. Since the system is
difficult to solve, we firstly choose the potential proportional to the square of scalar field and then we use the separation of variables techniques. Therefore, we obtain the non-trivial solution for the above set of differential equations (\ref{n1})-(\ref{n4}) as the follows
\begin{equation}\label{s1}
\alpha=-\frac{2\alpha_{0}}{3}a^{n+1}, \quad \beta=\alpha_{0}a^{n}\phi,
\end{equation}
\begin{eqnarray}\label{s2}
f(\phi)&=&-\frac{3}{8}\phi^{2}-\frac{3c_{1}}{2}\phi^{\frac{2(n+3)}{3}}, \nonumber \\
g(\phi)&=& \frac{1}{4}\phi^{2}+\frac{3c_{1}}{2(n+3)}\phi^{\frac{2(n+3)}{3}}+c_{2},
\end{eqnarray}
\begin{equation}\label{s3}
V(\phi)=\lambda \phi^2,
\end{equation}
where $c_{1}$, $c_{2}$, $\alpha_{0}$, $\lambda$ and $n$ are integration constants  and $n\neq-3$. From the values of symmetry
generator coefficients (\ref{s1}), the Noether symmetry generator is given by
\begin{equation}\label{vec1}
\textbf{X}=-\frac{2\alpha_{0}}{3}a^{n+1}\frac{\partial }{\partial a}+\alpha_{0}a^{n}\phi\frac{\partial}{\partial \phi}.
\end{equation}
For the special case $n=-3$, the Noether symmetry equations (\ref{n1})-(\ref{n4}) give the following solutions
\begin{equation}\label{s11}
\alpha=-\frac{2\alpha_{0}}{3}a^{-2}, \quad \beta=\alpha_{0}a^{-3}\phi,
\end{equation}
\begin{eqnarray}\label{s22}
f(\phi)&=&-\frac{3}{8}\phi^{2}+\frac{3c_{3}}{\phi^{2}}-\frac{3c_{1}}{2}, \nonumber \\
g(\phi)&=& \frac{1}{4}\phi^{2}+c_{1}ln(\phi)+c_{2},
\end{eqnarray}
with same potential function given by Eq. (\ref{s3}).

In this part we attempt to solve the basic cosmological equations of the scalar tensor teleparallel gravity model with a boundary term analytically. In order to integrate the dynamical systems (\ref{dec}), (\ref{fried}) and (\ref{kg}), we search for a cyclic variable associated with the Noether symmetry generator (\ref{vec1}). So, we introduce two arbitrary functions $z$ and $u$ defined as $z=z(a,\phi)$ and $u=u(a,\phi)$ respectively. The transformed Lagrangian is cyclic in one of the new variables so that the Lagrangian depending on new variables produces a reduced dynamical system which is generally solvable. Utilizing the relations $i_{\bf X} dz =1$ and $i_{\bf X} du =0$ where  $i_{\bf X}$ is the interior product operator of ${\bf X}$, we obtain the differential equations
\begin{equation}\label{c1}
\alpha\frac{\partial z}{\partial \dot{a}}+\beta\frac{\partial z}{\partial \dot{\phi}}=1,
\end{equation}
\begin{equation}\label{c2}
\alpha\frac{\partial u}{\partial \dot{a}}+\beta\frac{\partial u}{\partial \dot{\phi}}=0,
\end{equation}
respectively. A general discussion of this issue could be found in \cite{rit1,rit2,cap1,cap2}. Inserting the values of $\alpha$ and $\beta$ given by (\ref{n1}) into the equations (\ref{c1}) and (\ref{c2}), we obtain the following solutions
\begin{equation}\label{c3}
z=\frac{3}{2n\alpha_{0}}a^{-n},\quad u=a^{\frac{3}{2}}\phi,
\end{equation}
where $n\neq0$. For the case of $n=0$, the coupling function $f(\phi)$ is proportional to square of the scalar field and we take $c_{1}=-1/2$, then the coupling function $g(\phi)$ vanishes. Thus the action (\ref{action1}) reduces to the telleparalel dark energy model that is in depth analysed using the Noether symmetry approach in our previous work \cite{kucuk}. Correspondingly, the scale factor and scalar field could be expressed as
\begin{equation}\label{c4}
a=(\frac{2n\alpha_{0}}{3}z)^{-\frac{1}{n}},\quad \phi=u\left(\frac{2n\alpha_{0}}{3}z\right)^{\frac{3}{2n}}.
\end{equation}
Under this transformation, considering the coupling functions (\ref{s2}) and the potential (\ref{s3}) the Lagrangian (\ref{lag}) in terms of new variables takes the suitable form
\begin{equation}\label{nl}
\mathcal{L}=\frac{\dot{u}^2}{2}-2c_{1}\alpha_{0}u^{\frac{2n+3}{3}}\dot{u}\dot{z}-\lambda u^2,
\end{equation}
in which one can easily see that $z$ is a cyclic variable.
The new Lagrangian provide the following equations of motion:
\begin{equation}\label{r1}
-2c_{1}\alpha_{0}u^{\frac{2n+3}{3}}\dot{u}=I_{0},
\end{equation}
\begin{equation}\label{r2}
\ddot{u}-2c_{1}\alpha_{0}u^{\frac{2n+3}{3}}\ddot{z}+2\lambda u=0,
\end{equation}
\begin{equation}\label{r3}
\frac{\dot{u}^2}{2}-2c_{1}\alpha_{0}u^{\frac{2n+3}{3}}\dot{u}\dot{z}+\lambda u^2=0,
\end{equation}
where $I_{0}$ is a constant of motion. The equation (\ref{r1}) can be easily integrated to give
\begin{equation}\label{sol1}
u(t)=\left(-\frac{I_{0}(n+3)}{3\alpha_{0}c_{1}}t+b_{1}\right)^{\frac{3}{2(n+3)}},
\end{equation}
where $b_{1}$ is an arbitrary constant of integration and $n\neq-3$. Firstly, we consider the case $n=-3$. For this case, the Noether symmetry approach yields the coupling functions and potential given by the Eqs. (\ref{s22}) and (\ref{s3}) with the symmetry generator coefficients (\ref{s11}). Now, we can easily solve Eqs. (\ref{r1})-(\ref{r3}) for $z$ and $u$. Utilizing the obtained solution for $z$ and Eq. (\ref{c4}), we find the scale factor as $a(t)^3=a_{0}e^{-\frac{I_{0}t}{2c_{1}\alpha_{0}}}+c_{2}$ where $a_{0}$ and $c_{2}$ are constants being combinations of other constants. This solution for $c_{2}=0$ gives us the de Sitter Universe.
Secondly in general case, by inserting the solution (\ref{sol1}) into Eqs. (\ref{r2}) and (\ref{r3}), we obtain the following solution for $z$,
\begin{eqnarray}\label{sol2}
z(t)=\left[\frac{8\lambda n\left(I_{0}(n+3)t-3\alpha_{0}c_{1}b_{1}\right)^{2}-9I_{0}^{2}(n+6)}{24I_{0}^2\alpha_{0}c_{1}n(n+6)}\right]\nonumber \\ \times\left(\ell t+b_{1}\right)^{-\frac{n}{n+3}}+b_{2},
\end{eqnarray}
where $b_{1}$ is an another constant of integration, we define $\ell=-\frac{I_{0}(n+3)}{3\alpha_{0}c_{1}}$ and $n\neq-6$ and $I_{0}\neq0$. Therefore, the exact solution of the scale factor and scalar field  could be given out as below
\begin{eqnarray}\label{scale}
a(t)=\left[\frac{8\lambda n\left(I_{0}(n+3)t-3\alpha_{0}c_{1}b_{1}\right)^{2}-9I_{0}^{2}(n+6)}{36I_{0}^2c_{1}(n+6)}\right]^{-\frac{1}{n}} \nonumber \\ \times
\left(\ell t+b_{1}\right)^{\frac{1}{n+3}},
\end{eqnarray}
\begin{eqnarray}\label{scalar}
\phi(t)=\left[\frac{8\lambda n\left(I_{0}(n+3)t-3\alpha_{0}c_{1}b_{1}\right)^{2}-9I_{0}^{2}(n+6)}{36I_{0}^2c_{1}(n+6)}\right]^{\frac{3}{2n}} \nonumber \\ \times
\left(\ell t+b_{1}\right)^{-\frac{3}{n(n+3)}},
\end{eqnarray}
where we take that $b_{2}$ is zero without loss of generality. It is very difficult to analyze the solution (\ref{scale}) in this form. Hence, we follow
the procedure used in Ref.\cite{g1,g2} by setting the present time $t_{0}=1$. (see also detailed this procedure()) We assume first that at $t=0$, $a(0)=0$ so that we fix the origin of time. This condition gives, with the use of scale factor (\ref{scale}), $b_{1}=0$ for $n>0$ or $8\lambda n (\alpha_{0}c_{1}b_{1})^{2}-I_{0}^2(n+6)=0$, $b_{1}\neq0$ for $n<0$. At this point we restrict to ourselves to the case $b_{1}=0$ for $n>0$. The second condition is to set $a_{0}\equiv a(t_{0}=1)=1$ which is standard, and finally the Hubble constant are constrained $H(t_{0}=1)\equiv \mathcal{H}_{0}$. Because of this choice of time unit, it turns out that our $\mathcal{H}_{0}$ is not the same as the $H_{0}$ that appears in the standard FRW model. By means of these choices, we obtain the scale factor and the Hubble parameter as follows
\begin{equation}\label{scale2}
a(t)=\left[\frac{c_{1}m-n\left[\mathcal{H}_{0}(n+3)-1\right]t^2}{2(n+3)}\right]^{-\frac{1}{n}}t^{\frac{1}{n+3}},
\end{equation}
\begin{equation}\label{hubble}
H(t)=\frac{m+(n+6)\left[\mathcal{H}_{0}(n+3)-1\right]t^2}{(n+3)\left[mt-n\left(\mathcal{H}_{0}(n+3)-1\right)t^3\right]},
\end{equation}
where we define $m=n\left[\mathcal{H}_{0}(n+3)+1\right]+6$. The deceleration parameter which is defined by $q=-\frac{\ddot{a}a}{\dot{a}^2}$ is useful to study current expansion of the universe. So the universe expands in an accelerated behavior for $q < 0$ while $q > 0$ means a decelerated expansion of the universe. For our model, this parameter turns out to be
\begin{eqnarray}\label{dece}
q=-1- \frac{n(n+3)(n+6)\left[\mathcal{H}_{0}(n+3)-1\right]^2t^4}{\left[m+(n+6)
\left[\mathcal{H}_{0}(n+3)-1\right]t^2\right]^2} \nonumber \\ +\frac{(n+3)\left[m^2-2m(2n+3)\left[\mathcal{H}_{0}(n+3)-1\right]t^2\right]}{\left[m+(n+6)
\left[\mathcal{H}_{0}(n+3)-1\right]t^2\right]^2}.
\end{eqnarray}
On the other hand, using the energy density (\ref{ed}) and pressure (\ref{pres}) of the scalar field, the equation of state parameter takes the following form in the model
\begin{eqnarray}\label{omega}
\omega_{\phi}=-1 - \frac{2 n (n+3)(n+6)\left[\mathcal{H}_{0}(n+3)-1\right]^2t^4}{3\left[m+(n+6)
\left[\mathcal{H}_{0}(n+3)-1\right]t^2\right]^2} \nonumber \\ +\frac{2(n+3)\left[m^2-2m(2n+3)\left[\mathcal{H}_{0}(n+3)-1\right]t^2\right]}{3\left[m+(n+6)
\left[\mathcal{H}_{0}(n+3)-1\right]t^2\right]^2}.
\end{eqnarray}
\begin{figure}
\resizebox{0.48\textwidth}{!}{\includegraphics{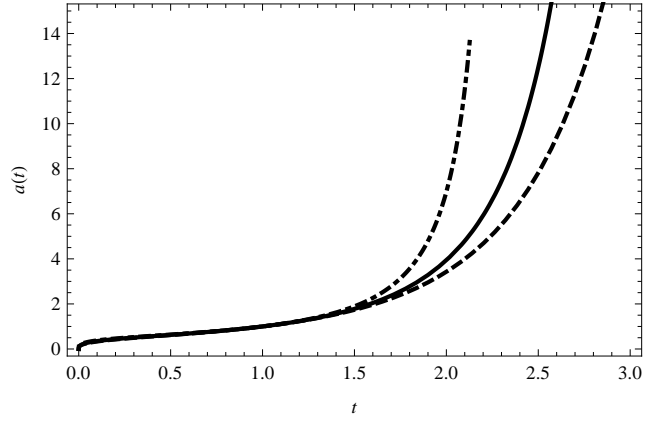}}
\caption{Plots of the scale factor with respect
to cosmic time $t$ for the different value of $n$. The dashed
line is for the value $n=0.0001$, the solid line is for the value $n=0.2$ and the
dot dashed line is for the value $n=0.6$. We take $\mathcal{H}_{0}=1$ and $c_{1}=-1$.} \label{fig:1}
\end{figure}
\begin{figure}
\resizebox{0.48\textwidth}{!}{\includegraphics{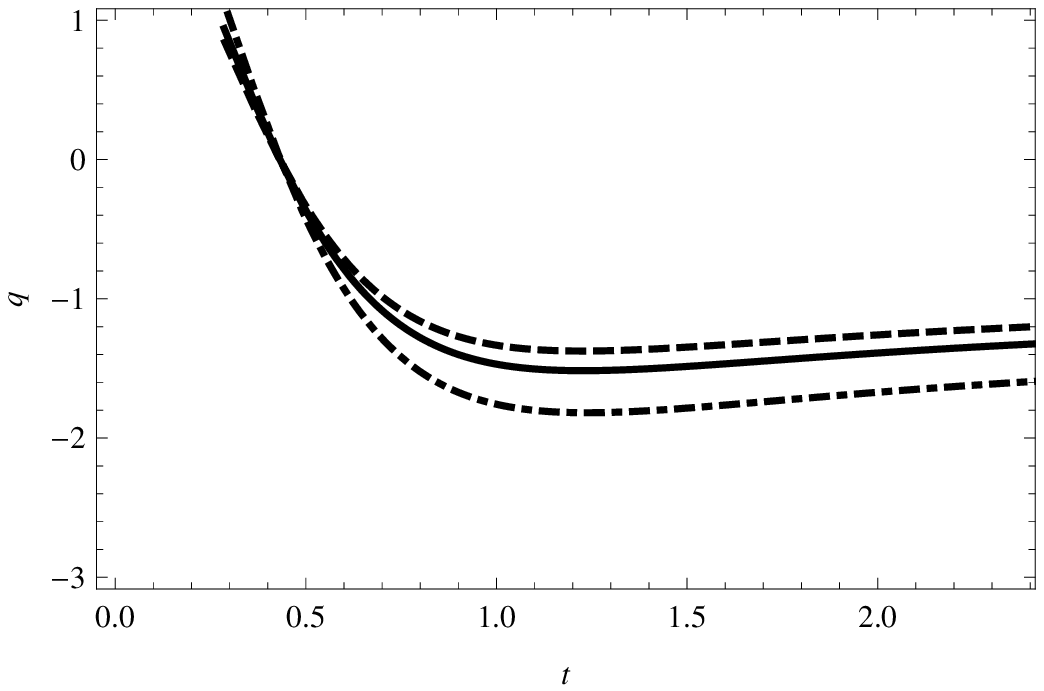}}
\caption{Plots of the deceleration parameter with respect
to cosmic time $t$ for the different value of $n$. The dashed
line is for the value $n=0.0001$, the solid line is for the value $n=0.2$ and the
dot dashed line is for the value $n=0.6$. We take $\mathcal{H}_{0}=1$.} \label{fig:2}
\end{figure}
\begin{figure}
\resizebox{0.48\textwidth}{!}{\includegraphics{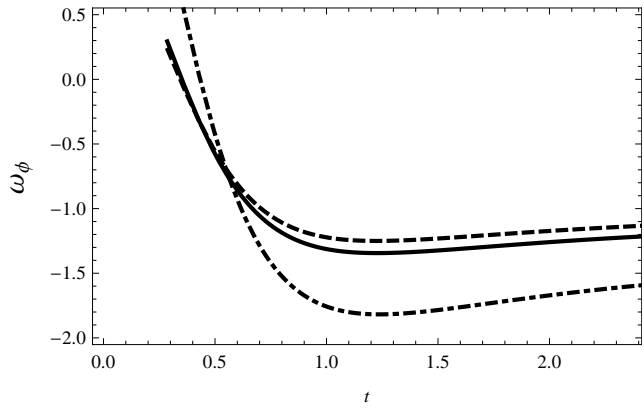}}
\caption{Plots of the equation of state parameter with respect
to cosmic time $t$ for the different value of $n$. The dashed
line is for the value $n=0.0001$. The solid line is for the value $n=0.2$ and the
dot dashed line is for the value $n=0.6$. We take $\mathcal{H}_{0}=1$.} \label{fig:3}
\end{figure}
The graphical analysis of the scale factor versus cosmic time $t$ for the different value of $n$ represented in the Figure (\ref{fig:1}). As one can see from this figure, the universe describes an expansionary phase with the scale factor a monotonically increasing function of time. The deceleration parameter, depicted in Figure (\ref{fig:2}), indicates the existence of a transition from a decelerating phase to an accelerating one. Figure (\ref{fig:3}) shows behavior of the equation of state parameter with respect to cosmic time $t$ for the different value of $n$. From this figure, we observe that crossing of the phantom divide line $\omega_{\phi}=-1$ can be realised from the quintessence phase $\omega_{\phi}>-1$ to phantom phase $\omega_{\phi}<-1$ in our model described by the Noether symmetry solution. We also note that both cosmological parameters go to the value $-1$ in the large time limit for small values of the parameter $n$.

Now, we return to the solution of field equations (\ref{r1})-(\ref{r1}) in the specially case of the constant of motion $I_{0}=0$. It can be seen below that this case has an interesting solution. If $I_{0}=0$, then from Eq. (\ref{r1}) we get a solution as $u(t)=u_{0}$ where $u_{0}$ is a constant. This solution satisfies Eq. (\ref{r3}) if $\lambda=0$ which gives a scalar-tensor teleparallel model with boundary term but without scalar potential. From Eq. (\ref{r2}) the variable $z(t)$ is solved as $z(t)=z_{0}t+z_{1}$ where $z_{0}$ and $z_{1}$ are an integration constant. Therefore, for $I_{0}=0$ the scale factor and scalar field are obtained as follows
\begin{equation}\label{scal2}
a(t)=\left[\frac{2n\alpha_{0}}{3}(z_{0}t+z_{1})\right]^{-\frac{1}{n}},
\end{equation}
\begin{equation}\label{scalr2}
\phi(t)=u_{0}\left[\frac{2n\alpha_{0}}{3}(z_{0}t+z_{1})\right]^{\frac{3}{2}}.
\end{equation}
From these considerations, it is easy to realize that any power law solution can be achieved according
to the value of $n$. For example, a pressureless matter solution is recovered for $a(t)\sim t^{\frac{2}{3}}$ with $n=-\frac{3}{2}$, a radiation solution is for $a(t)\sim t^{\frac{1}{2}}$ with $n=-2$. The deceleration parameter takes the form as follows
\begin{equation}\label{dece2}
q=-1-n,
\end{equation}
which means that
The equation of state parameter become
\begin{equation}\label{eos2}
\omega_{\phi}=-1-\frac{2n}{3}.
\end{equation}
From Eqs. (\ref{dece2}) and (\ref{eos2}), if $-1<n<0$, then we have $q<0$ and $-1<\omega_{\phi}<-\frac{1}{3}$ which corresponds to a universe with quintessence phase. So that the universe is both expanding and accelerating in this case. if $n<-1$, then we obtain $q>0$ and $\omega_{\phi}>-\frac{1}{3}$ which corresponds to a universe with decelerating expansion. On the other hand, if $n>0$, then we have $q<0$ and $\omega_{\phi}<-1$ which corresponds to a universe with phantom phase. So that the universe is accelerating but shrinking in this case. We also note that the limit of $n\rightarrow 0$ in Eq. (\ref{eos2}) corresponds to the limit of $\omega_{\phi}\rightarrow -1$, which is consistent with the $\Lambda$CDM.
model.

\section{Summary and Conclusion} \label{sec:3}

There exist some methods to investigate the integrability of a dynamical system. In this study, we chose to find Noether symmetries of the point-like Lagrangian of a scalar tensor teleparallel gravity theory to obtain the conserved quantities. For that gravitational Lagrangian, we considered a model including the scalar field which is nonminimal coupled to the torsion and boundary term where the boundary term represents the divergence of the torsion vector. This model is important since it shows some interesting aspects in cosmology and in describing the late time acceleration of the Universe. Furthermore, the model is reduced to the theories such as quintessence, teleparallel dark energy and non-minimally coupled scalar field to the Ricci scalar under the suitable limits.

As above mentioned, the Noether symmetry approach is important because it can be considered as a physically motivated criterion so that such a symmetry are always related to conserved quantities. The existence of Noether symmetry also restricts the forms of the unknown functions in a given Lagrangian (i.e. in particular
coupling functions and scalar potential), and allows us to find a transformation given by (\ref{c4}) in which the scale factor and the scalar field are
written in terms of new dynamical variables where one of the variables is cyclic. Under these transformations, Lagrangian is reduced to a simpler form. In this study, the equations of motion of the considered model for the FRW space-time background in the form of equations (\ref{dec})-(\ref{kg}) have obtained. By applying the Noether symmetry approach, we have found the explicit forms of the coupling functions $f(\phi)$ and $g(\phi)$ and potential of the scalar field $V(\phi)$ as Eqs. (\ref{s2}) and (\ref{s3}) respectively. By introducing cyclic variables, we have found some exact cosmological solutions of the corresponding field equations using these forms obtained by the existence of Noether symmetry. The properties of the cosmological parameters relevant to the solutions have been analyzed in detail. The main, and interesting feature of these solutions is that they describe an accelerating expansion of the universe. We have also observed the equation of state parameter shows that crossing of the phantom divided line can be realized (see Fig. (\ref{fig:3})). Therefore, it is important to investigate Noether symmetries of the teleparallel dark energy models with the boundary term to explain the late time acceleration of the universe.

%\section*{Acknowledgements}
\begin{acknowledgements}
This work was supported by Akdeniz University, Scientific Research Projects Unit.
\end{acknowledgements}

%\begin{acknowledgements}
%\end{acknowledgements}



\end{document}